\documentclass[conference,a4paper]{APSIPA2020}
\usepackage{multirow}
\usepackage{graphicx}
\usepackage{amsmath}
\usepackage{amssymb}
\usepackage{amsxtra}
\usepackage{threeparttable}

\usepackage{cite}
\usepackage{multicol, multirow, booktabs}

\usepackage{amsmath,amsfonts,amssymb,bm}

\newcommand{\ie}{\textit{i.e.}}
\newcommand{\eg}{\textit{e.g.}}

\newcommand{\etal}{\textit{et~al.}}

\newcommand{\mB}{\mathbf{B}}

\newcommand{\mF}{\mathbf{F}}
\newcommand{\mG}{\mathbf{G}}

\newcommand{\mX}{\mathbf{X}}
\newcommand{\mY}{\mathbf{Y}}

\fussy

\begin{document}

\title{Tatum-Level Drum Transcription
Based on a Convolutional Recurrent Neural Network 
with Language Model-Based Regularized Training}

\author{
\authorblockN{
Ryoto Ishizuka,
Ryo Nishikimi,
Eita Nakamura, and 
Kazuyoshi Yoshii}
\authorblockA{
Graduate School of Informatics, Kyoto University, Kyoto, Japan \\
E-mail: \{ishizuka, nishikimi, enakamura, yoshii\}@sap.ist.i.kyoto-u.ac.jp
}
}

\maketitle
\thispagestyle{empty}

\begin{abstract}
This paper describes a neural drum transcription method
 that detects from music signals
 the onset times of drums
 at the \textit{tatum} level,
 where tatum times are assumed to be estimated in advance.
In conventional studies on drum transcription,
 deep neural networks (DNNs) have often been used
 to take a music spectrogram as input 
 and estimate the onset times of drums at the \textit{frame} level.  
The major problem with such frame-to-frame DNNs, however, is that
 the estimated onset times do not often conform with
 the typical tatum-level patterns appearing in symbolic drum scores
 because the long-term musically meaningful structures of those patterns 
 are difficult to learn at the frame level.
To solve this problem,
 we propose a regularized training method for a frame-to-tatum DNN.
In the proposed method,
 a tatum-level probabilistic language model 
 (gated recurrent unit (GRU) network or repetition-aware bi-gram model) 
 is trained from an extensive collection of drum scores.
Given that the musical naturalness of tatum-level onset times
 can be evaluated by the language model,
 the frame-to-tatum DNN is trained 
 with a regularizer based on the pretrained language model. 
The experimental results demonstrate
 the effectiveness of the proposed regularized training method.
\end{abstract}

\section{Introduction}
Automatic drum transcription (ADT) 
 is a challenging subtask in automatic music transcription (AMT)
 that aims to estimate \textit{symbolic} musical scores 
 from music signals.
This is an important task
 because the drum part forms the rhythmic backbone of popular music.
In this paper,
 we focus on the three main drum instruments of a drum kit:
 bass drums (BD), snare drums (SD), and hi-hats (HH).
In general, the estimated onset times of drums
 are represented at the frame level (in seconds)
 and very few ADT methods
 aim to estimate symbolic drum scores on regular time grids.

\begin{figure}[t]
\centering
\includegraphics[width=.9\columnwidth]{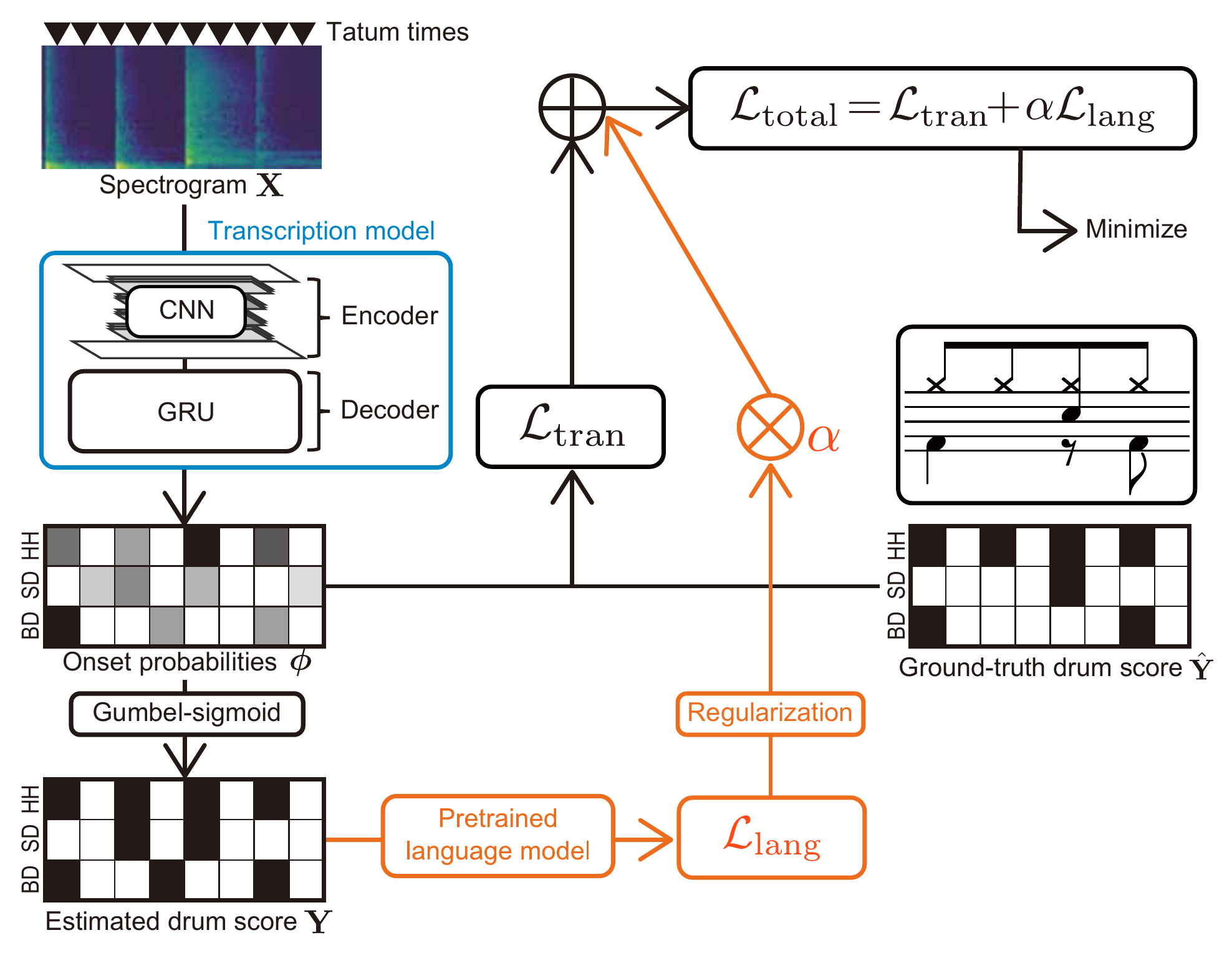}
\vspace{-1mm}
\caption{
Supervised training of a neural drum transcription model
with musical naturalness-aware output regularization 
based on a pretrained language model.}
\label{fig:overview}
\vspace{-4mm}
\end{figure}

In ADT, deep learning
 and nonnegative matrix factorization (NMF)
 have been two major approaches 
 to estimating the onset activations of drums
 from audio spectrograms \textit{at the frame level}~\cite{Wu2018review}.
In particular, 
 convolutional neural networks (CNNs)
 that can extract useful features 
 from local time-frequency regions
 have shown good performances
 \cite{jacques2018automatic, gajhede2016convolutional, southall2017automatic}.
Recurrent neural networks (RNNs) have also been used
 for learning the frame-level long-term dependency 
 of onset activations\cite{vogl2016recurrent, stables2016automatic,vogl2017drum}.
Because the spectrogram of a drum part 
 consists of a number of repetitions of the same impulsive sounds,
 NMF has still been used extensively for ADT~\cite{wu2015drum, roebel2015automatic, ueda2019bayesian, dittmar2014real, paulus2009drum}.

However,
 as these purely frame-level ADT methods
 have no mechanism to prevent the estimated onset times of drums
 from having a musically unnatural structure,
 the rhythmic and repetitive patterns of drum onsets
 are hard to be learned at the frame level.
One solution to this problem is
 to use a \textit{language model} (LM)
 that represents a probability distribution of drum onsets \textit{at the tatum level}
 such that the musical naturalness of drum scores can be evaluated.
Such an LM (drum score prior)
 has been integrated with an NMF-based acoustic model (drum score likelihood)
 in a Bayesian manner \cite{ueda2019bayesian}.
The performance of this method, however, remains unsatisfactory
 because of the limited expressive power of NMF
 and the time-consuming posterior inference of drum scores
 required at run-time.

LMs play an essential role in automatic speech recognition (ASR)
 for estimating a word sequence from a feature sequence
 such that the estimated word sequence 
 is syntactically and semantically coherent.
The classical yet effective approach to ASR
 is to combine a \textit{word-level} language model (\eg, n-gram model) 
 representing the generative process of a word sequence
 with a \textit{frame-level} acoustic model (\eg, hidden Markov model (HMM))
 representing the generative process of a feature sequence from a word sequence~\cite{hinton2012deep}.
To infer a word sequence from a feature sequence using Bayes' theorem,
 a sophisticated decoder (\eg, weighted finite-state transducer (WFST))
 based on the language and acoustic models is used at run-time.
The advantage of this approach lies in its modularity;
 the language and acoustic models can be trained
 from text data and paired data (speech data with transcriptions), respectively.

Recently, the end-to-end approach to ASR has been actively investigated
 for directly inferring a word sequence
 from a feature sequence with a deep neural network (DNN).
A popular choice is 
 to use an encoder-decoder architecture with an attention mechanism,
 where the encoder and decoder are conceptually considered 
 to have acoustic and language modeling capabilities, respectively~\cite{chorowski2014end}.
While such a network is easy to implement and works fast at run-time,
 only paired data can be used for training the whole network,
 which means that massive text data cannot be used
 for improving the language modeling capability of the network.
To use the knowledge of an LM
 trained on huge text data in an end-to-end recognizer,
 knowledge transfer techniques~\cite{hinton2015distilling} 
 have been investigated~\cite{bai2019learn, chen2019distilling}.

In light of these circumstances,
 we propose an ADT method based on a convolutional RNN (CRNN)
 that directly infers a sequence of \textit{tatum-level} onset times
 from a sequence of \textit{frame-level} mel spectra (Fig.~\ref{fig:overview}).
We do not directly use an encoder-decoder architecture with an attention mechanism,
 which has widely been used for sequence-to-sequence learning
 in various applications, including ASR.
In practice, the beat times can be estimated accurately 
 for typical popular music with regular rhythmic structure (our main target)
 and attention-based alignment between long frame- and tatum-level sequences 
 is hard to learn from a limited amount of training data.
We thus use the estimated tatum times instead of using an attention mechanism
 for combining the frame-level convolutional layers (encoder) 
 extracting useful features from mel spectra
 and tatum-level recurrent layers (decoder)
 learning the rhythmic and repetitive patterns of drum onsets.

To transfer the knowledge of an LM
 trained from a large number of drum scores,
 we train the CRNN in a regularized manner.
More specifically, we aim to minimize
 the weighted sum of the transcription error $\mathcal{L}_{\mathrm{tran}}$
 and the musical unnaturalness $\mathcal{L}_{\mathrm{lang}}$
 computed for the CRNN output,
 where $\mathcal{L}_{\mathrm{tran}}$ is the cross entropy
 between the estimated \textit{soft} drum score
 and the ground-truth score
 and $\mathcal{L}_{\mathrm{lang}}$ is the LM-based negative log-probability  
 of the estimated \textit{hard} (binarized) score.
Note that the hard score is obtained by applying the gumbel-sigmoid trick~\cite{tsai2018learning}
 to the soft score in a differentiable manner for backpropagation-based optimization.

\section{Related Work}
\label{sec:related work}

This section reviews related work
 on AMT and ADT based on language models
 and knowledge transfer.

\subsection{Automatic Drum Transcription}

Nonnegative matrix factorization (NMF)
 has often been used for decomposing a drum-part spectrogram
 into the spectra and temporal activations of drums~\cite{dittmar2014real, wu2015drum, paulus2009drum}.
To overcome the limited expressive power of NMF,
 CNNs have been used in ADT~\cite{gajhede2016convolutional, southall2017automatic, jacques2018automatic}
 for automatically extracting local features
 as well as in AMT~\cite{schluter2014improved}.
RNNs have also been proposed
 for capturing the long-term temporal dependency
 at the frame level.
Vogl~\etal introduced RNNs~\cite{vogl2016recurrent}
 as well as multi-task learning~\cite{vogl2017drum} in ADT.
In such ways,
 DNN-based transcription methods,
 which are trained by paired data consisting of audio signals with annotations,
 have achieved high performances.

\subsection{Language model}

One way of improving AMT and ADT methods
 is to introduce an LM
 that evaluates the musical naturalness of estimated scores.
Such LMs have generally been formulated at the frame level.
Raczy{\`n}ski~\etal~\cite{raczynski2013dynamic}, for example,
 used a deep belief network for modeling a transition of chord symbols
 and improved the chord recognition system consisting of NMF.
Sigtia \etal~\cite{sigtia2015audio} used an LM
 for estimating the most likely chord sequence
 from the chord posterior probabilities estimated by an RNN-based chord recognition system.
As pointed out in~\cite{korzeniowski2017futility,ycart2019blending}, 
 however, LMs can be more effectively formulated at the tatum level 
 for learning the musically meaningful structure.

Tatum-level LMs have recently been used for AMT and ADT.
Korzeniowski \etal~\cite{korzeniowski2018automatic}
 used N-gram for a symbolic LM
 and improved the DNN-based chord recognition system.
Korzeniowski \etal~\cite{korzeniowski2018improved}
 also insisted that a frame-level LM can only smooth the onset probabilities of chord symbols,
 and they used an RNN-based symbolic LM with a chord duration model.
Ycart \etal~\cite{ycart2017study}
 investigated the predictive power of LSTM networks
 and demonstrated that
 a long short-term memory (LSTM) working at the level of 16th note timesteps
 could express the musical structure
 such as note transitions.
Thompson \etal~\cite{thompson2014drum}
 used a template-based LM for classifying audio signals
 into a limited number of drum patterns
 with a support vector machine (SVM).
Ueda \etal~\cite{ueda2019bayesian}
 proposed a Bayesian approach 
 using a DNN-based LM as a prior of drum scores.
Integration of a tatum-level LM into a DNN-based ADT system, however,
 has still been an open problem.

\subsection{Knowledge Transfer}

Transfer learning aims 
 to effectively use knowledge 
 from a related domain~\cite{weiss2016survey},
 and has been used in various fields.
This method can be used for labeled data
 as well as unlabeled data.
In the student-teacher framework,
 some studies have attempted to train a student model
 that has the same capacity as a teacher model~\cite{mobahi2020self}
 for achieving higher accuracy~\cite{furlanello2018born}.
Transfer learning using a compact student model
 is often called knowledge distillation~\cite{hinton2015distilling}.

Recently,
 there have been some studies 
 on using knowledge learned from
 an extensive collection of unpaired data
 for improving other models.
In ASR,
 an LM was integrated into ASR systems
 to generate more syntactically or semantically word sequences.
These methods, however,
 require an LM in decoding
 and take much time in the inference stage.
More recently,
 an ASR system based on knowledge distillation was proposed,
 where an LM softened a probability distribution as a regularizer
 to transfer the knowledge of unpaired data.
The method did not require an LM in the inference stage~\cite{bai2019learn}.
The idea of knowledge distillation
 was also used in ADT~\cite{wu2017automatic},
 where an NMF-based teacher model was applied
 to a DNN-based student model,
 and this method showed great potential
 to utilize unpaired data.
Note that in the transfer learning,
 the same or different datasets are used
 depending on the problem specification
 ~\cite{hinton2015distilling, zagoruyko2016paying, yim2017gift}.

\section{Proposed method}
\label{sec:proposed method}

This section describes the proposed ADT method
 that estimates a drum score
 from the mel spectrogram of a music signal
 (Section~\ref{sec:problem_specification}).
As shown in Fig.~\ref{fig:overview},
 our method uses a CRNN-based transcription model
 for estimating the onset probabilities of drums at the tatum level
 (Section~\ref{sec:transcription_model}).
Given that
 a pretrained LM of drum scores 
 can be used for evaluating the musical naturalness of a drum score
 (Section~\ref{sec:language_model}),
 the transcription model is trained in a supervised manner
 with a regularization mechanism 
 based on the pretrained language model 
 (Section~\ref{sec:regularized_training}).

\subsection{Problem Specification}
\label{sec:problem_specification}

Our goal is to estimate a drum score $\mY \in \{0, 1\}^{K \times M}$
 from the mel spectrogram of a target musical piece
 $\mX \in {\mathbb R}_+^{F \times T}$,
 where $K$ is the number of drum instruments (BD, SD, and HH, \ie, $K=3$),
 $M$ the number of tatums,
 $F$ the number of frequency bins,
 $T$ the number of time frames. 
In this paper,
 we assume that all onset times are located on the tatum-level
 (quarter-beat-level) grid
 and the tatum times $\mB = \{b_m\}_{m=1}^M$ are estimated in advance.

\subsection{Transcription Model}
\label{sec:transcription_model}

The transcription model is used for estimating
 the tatum-level onset probabilities
 $\bm\phi \in [0, 1]^{K \times M}$,
 where $\phi_{k, m}$ represents the posterior probability 
 that drum $k$ has an onset at tatum $m$. 
The estimated drum score $\mY$ can be obtained
 by binarizing $\bm\phi$ with a threshold $\delta \in [0, 1]$.
The transcription model is implemented
 as a CRNN consisting of a frame-level encoder based on convolutional layers
 and a tatum-level decoder based on GRU layers followed by a fully-connected layer (Fig.~\ref{fig:acoustic_model}).
The encoder
 converts the mel spectrogram $\mX$ 
 to the latent features $\mF \in {\mathbb R}^{D \times T}$,
 where $D$ is the feature dimension.
The frame-level features $\mF$
 are then summarized to the tatum-level features $\mG \in {\mathbb R}^{D \times M}$
 through a max pooling layer referring to the tatum times $\mB$ as follows:
\begin{align}
G_{d,m} = \max_{\frac{b_{m-1}+b_m}{2} \le t < \frac{b_{m}+b_{m+1}}{2}} F_{d,t},
\end{align}
where $b_0=b_1$ and $b_{M+1}=b_{M}$. 
The decoder finally converts $\mG$ to the onset probabilities $\bm\phi$
 while considering the temporal dynamics of drum scores. 

\subsection{Language Model}
\label{sec:language_model}

The LM is used for estimating the generative probability (musical naturalness) of a drum score.
To achieve this, using an arbitrary existing drum score $\tilde\mY$\footnote{
 For brevity, we assume that only one drum score is used as training data.
 In practice, a sufficient number of drum scores are used.},
 the LM should be trained beforehand in an unsupervised manner
 such that the following negative log-likelihood for $\tilde\mY$ is minimized:
\begin{align}
\mathcal{L}_{\mathrm{lang}}(\tilde\mY)
= 
- \sum_{m=1}^{M} \log p(\tilde{Y}_{:,m} | \tilde{Y}_{:,1:m-1}),
\label{sec:l_lang}
\end{align}
where ``$i{:}j$'' indicates a set of indices from $i$ to $j$
 and ``$:$'' indicates all possible indices.
In this paper, we propose two LMs:
a \textit{skip-type} bi-gram model and a neural language model.

\subsubsection{Repetition-Aware Bi-Gram Model}

One possibility is to use a naive yet effective bi-gram model.
Assuming that
 popular music tends to have the 4/4 time signature
 and the same drum patterns tend to be repeated for making the rhythmic backbone,
 we propose a \textit{skip-type} bi-gram model
 representing the bar-level repetitive structure of $\tilde\mY$ as follows:
\begin{align}
 p(\tilde{Y}_{:,m} | \tilde{Y}_{:,1:m-1})
 &=
 \prod_{k=1}^K p(\tilde{Y}_{k,m} | \tilde{Y}_{k,m-16})
 \nonumber\\
 &=
 \prod_{k=1}^K \pi_{\tilde{Y}_{k,m-16},\tilde{Y}_{k,m}},
\end{align}
where $\pi_{A,B} \ (A, B \in \{0,1\})$ indicates the transition probability from $A$ to $B$.
Note that this model assumes the independence of the $K$ drums.

\subsubsection{Gated Recurrent Unit Model}

Another possibility is
 to use a more powerful neural LM based on GRUs 
 for directly representing $p(\tilde{Y}_{:,m} | \tilde{Y}_{:,1:m-1})$
 without assuming the independence of the $K$ drums.
This model is expected to implicitly represent different time signatures
 and consider a longer-range musically-meaningful temporal structure of drum scores.

\begin{figure}[t]
\centerline{
\includegraphics[width=.9\columnwidth]{
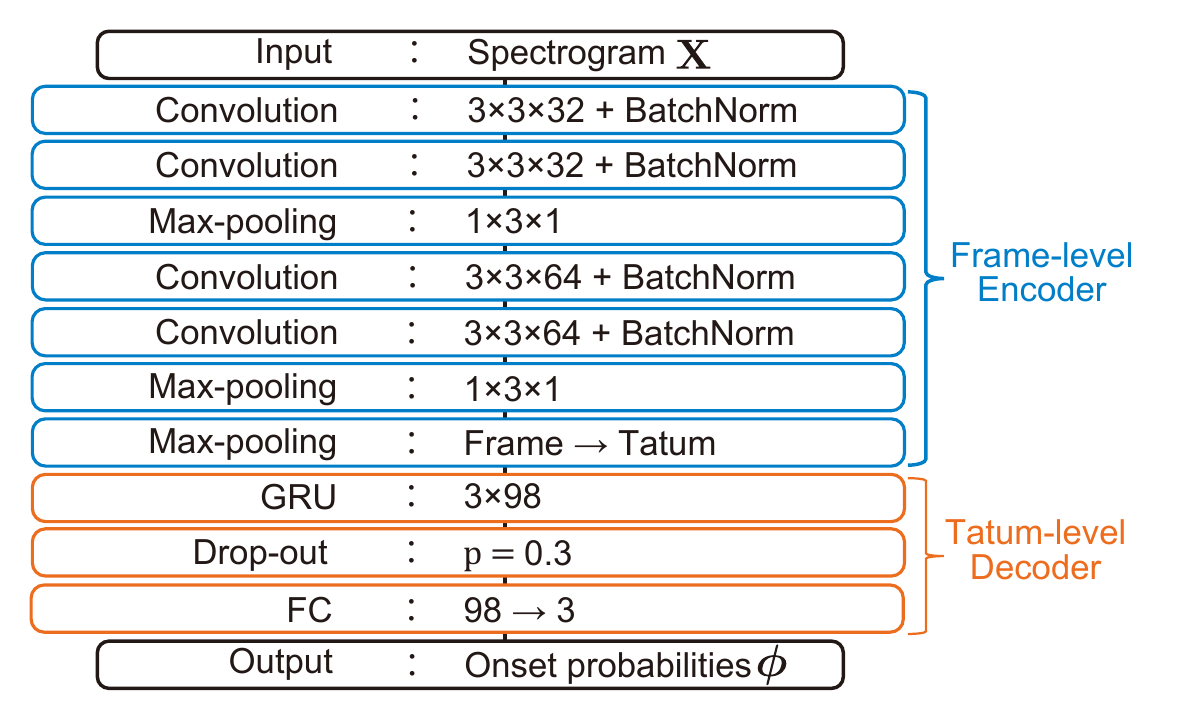}}
\vspace{-2mm}
\caption{
Configuration of the transcription model.
}
\vspace{-2mm}
\label{fig:acoustic_model}
\end{figure}

\subsection{Regularized Training}
\label{sec:regularized_training}

Given a ground-truth score $\hat\mY$,
 one can train the transcription model in a supervised manner
 such that the following modified negative log-likelihood for $\hat\mY$ is minimized:
\begin{align}
&\mathcal{L}_{\mathrm{tran}}(\bm\phi|\hat\mY)
\nonumber\\
&=
- \sum_{k=1}^{K} \sum_{m=1}^{M}
\bigl(\gamma \hat{Y}_{k, m}\log \phi_{k, m} 
\!+\! 
(1 \! - \! \hat{Y}_{k, m}) \log (1 \! - \! \phi_{k, m}) \bigr),
\label{sec:l_tran}
\end{align}
where
 $\gamma > 0$ is a weighting factor 
 compensating for the imbalance of the numbers of onset and non-onset tatums.
Because ${\mathcal L}_{\mathrm{tran}}$ evaluates only the transcription incorrectness 
 (cross entropy between $\bm\phi$ and $\hat\mY$),
 the musical naturalness of the estimated score $\mY$ obtained by binarizing $\bm\phi$
 is not considered.

To solve this problem, we propose an LM-based regularized training method
 that minimizes
\begin{align}
{\mathcal L}_{\mathrm{total}}
&=
{\mathcal L}_{\mathrm{tran}}(\bm\phi|\hat\mY)
+ 
\alpha {\mathcal L}_{\mathrm{lang}}(\bm\mY),
\label{sec:l_total}
\end{align}
where 
 $\alpha > 0$ is a weighting factor.
To use a backpropagation technique for optimizing the transcription model,
 the binary score $\mY$ should be obtained 
 from the soft representation $\bm\phi$ in a differentiable manner
 instead of simply binarizing $\bm\phi$ with a threshold.
We thus use a differentiable sampler
 called the gumbel-sigmoid trick~\cite{tsai2018learning} as follows:
\begin{align}
    U_{k, m}^{(i)} &\sim
    {\rm Uniform}(0,1),
    \\
    V_{k, m}^{(i)} &=
    -\log \left\{
    -\log\left( U_{k, m}^{(i)} \right)
    \right\},
    \\
    Y_{k, m} &=
    \sigma \left\{
    \frac{\phi_{k, m}+V_{k, m}^{(1)}-V_{k, m}^{(2)}}{\tau}
    \right\},
\end{align}
where
 $i=1,2$, $\tau > 0$ is a temperature,
 and $\sigma(\cdot)$ is a sigmoid function ($\tau=0.2$ in this paper).
Note that the pretrained LM (bi-gram or GRU model) is used as a fixed regularizer in the training phase
 and is not used in the prediction phase.

\section{Evaluation}
\label{sec:evaluation}

This section reports experiments conducted 
 for validating the proposed LM-based regularized training of the neural transcription model for ADT.

\subsection{Experimental Conditions}

The RWC Popular Music Database~\cite{goto2002rwc} was used for evaluation.
Among 89 songs having drum parts,
 we used 65 songs with correct ground-truth annotations.
These songs were randomly split into training and testing data 
 for 3-fold cross validation,
 where 15$\%$ of the training data was taken
 as validation data.
To extract drum sounds from polyphonic music signals,
 we used a music separation method called Open-Unmix~\cite{fabian2019open}.
The music signals and the separated drum signals
 were used in the training phase
 and the separated drum signals were used in the prediction phase.
The spectrogram of
 each music signal sampled at 44.1kHz
 was obtained using short-time Fourier transform (STFT)
 with a Hann window of 2048 points (46 ms)
 and a shifting interval of 441 points (10 ms).
The mel-frequency spectrogram
 was calculated using a mel-filter bank with 80 bands from 20 Hz to 20,000 Hz.

To pretrain the LMs (bi-gram and GRU models described in Section~\ref{sec:language_model}),
 we used 512 external drum scores (Japanese popular songs and The Beatles),
 which have no overlap with the RWC Popular Music Database~\cite{goto2002rwc}.
The GRU model we used consisted of 3 GRU layers with 64 hidden dimensions,
 which were experimentally determined by a Bayesian optimization method called Optuna\cite{optuna_2019}
 via 3-fold cross validation with the 512 scores.
 
We used madmom~\cite{bock2016madmom} for beat estimation
 and the performance was measured using
 the precision rate $\mathcal{P}$,
 the recall rate $\mathcal{R}$, and
 the F-measure $\mathcal{F}$ given by
\begin{align}
  \mathcal{P} = \frac{N_c}{N_e},\quad
  \mathcal{R} = \frac{N_c}{N_g},\quad
  \mathcal{F} = \frac{2\mathcal{R}\mathcal{P}}{\mathcal{R} + \mathcal{P}}, \label{equation:F-measure}
  \end{align}
where $N_e$, $N_g$, and $N_c$ were
 the number of estimated beats,
 that of ground-truth beats, and
 that of correctly-estimated beats,
 respectively.
The estimated beat
 was judged as correct
 if it was within 50 ms
 from the ground-truth beat.
The mir\_eval library\cite{raffel2014mir_eval} was used 
 for computing $\mathcal{P}$, $\mathcal{R}$, and $\mathcal{F}$.

\begin{table}[t]
\centering
\caption{
The ratios of undetectable onset times in three groups.
}
\vspace{-2mm}
\begin{tabular}{ccc|ccc} \toprule
\multicolumn{3}{c|}{Madmom} & 
\multicolumn{3}{c}{Ground-truth} \\
\textit{conflict} & \textit{far} & \textit{conflict $\cup$ far} &
\textit{conflict} & \textit{far} & \textit{conflict $\cup$ far} \\ \midrule
0.43\% & 0.23\% & 0.65\% &
1.19\% & 0.29\% & 1.48\% \\ \bottomrule
\end{tabular}
\label{table:statistical_features}
\vspace{-2mm}
\end{table}

\subsection{Justification of Tatum-Level Transcription}

We validate the appropriateness of our tatum-level transcription approach
 because there are undetectable drum onsets
 if all the onset times of each drum 
 are assumed to be exclusively located on tatum (quarter-beat) times.
Such undetectable onsets are (doubly) categorized into two groups:
 \textit{conflict} and \textit{far}.
In our experiment, 
 to convert frame-level onset times (\eg, original ground-truth annotations) 
 into a tatum-level score (\eg, estimation target $\hat\mY$),
 each onset time was quantized to the closest tatum time.
If multiple onset times are quantized into the same tatum time,
 only one onset time can be detected,
 \ie, the other onset times are undetectable
 and categorized into the \textit{conflict} group.
If actual onset times are not within 50 ms from the closest tatum times,
 they are categorized into the \textit{far} group.

Table~\ref{table:statistical_features} shows
 the ratios of such undetectable onset times to the total number of actual onset times
 when the estimated or ground-truth beat times are used for quantization.
The beat tracking method~\cite{bock2016madmom} achieved 96.4\%
 for the 65 songs used for evaluation.
This result justifies our approach at least for the majority of typical popular music
 because the total ratio of undetectable onset times was sufficiently low.

\subsection{Evaluation of Language Modeling}
\label{subsection:evaluation_of_language_model}

We evaluated the performance of the pretrained LMs for the 65 songs.
The perplexities obtained by the skip-type bi-gram model and the GRU model 
 were 1.51 and 1.44, respectively (lower is better).
The predictive capability of the GRU model 
 was better than that of the bi-gram model
 because the bi-gram model assumed the 4/4 time signature with the simple repeating structure.
We also confirmed that
 the perplexities were much smaller
 than the chance rate of 2.
The LMs were expected to work as regularizers and
 guide the outputs of the transcription model
 into musically-natural drum patterns.

\begin{figure*}[t]
\centerline{
\includegraphics[width=\linewidth]{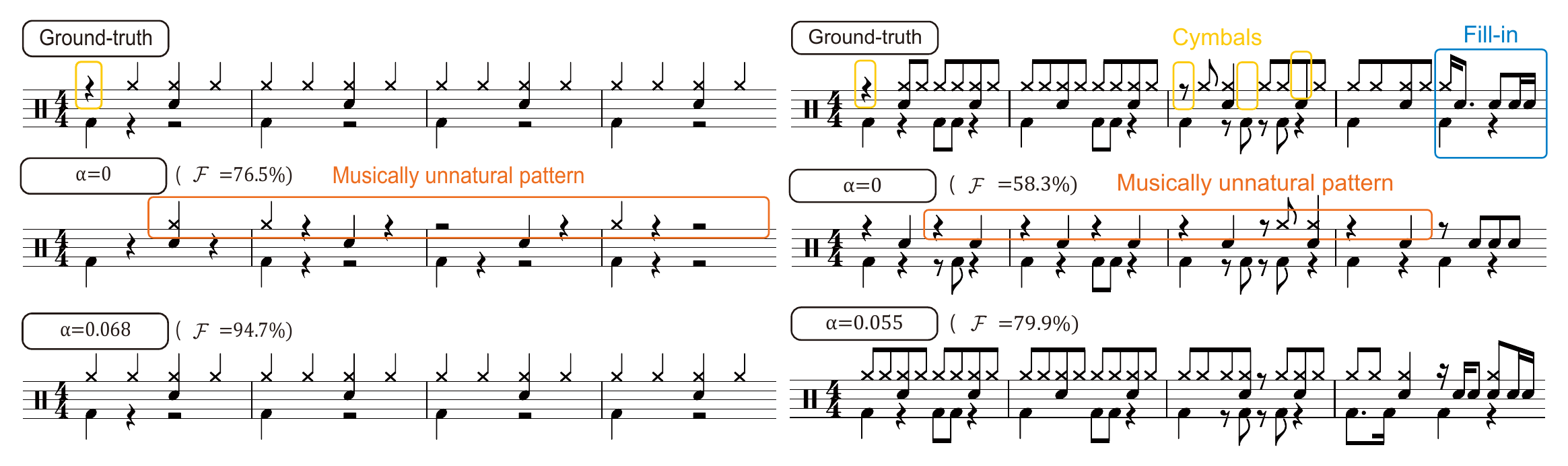}}
\vspace{-3mm}
\caption{
Examples of transcribed drum scores.
The left and right examples are excerpts from RWC-MDB-P-2001 No.~47 and No.~88, respectively.
The F-measure computed for each song was shown.
The top, middle, and bottom scores show the ground truth, 
 the estimated score obtained by the unconstrained transcription model,
 and the estimated score obtained by the transcription model
 regularized by the bi-gram model (left) and GRU model (right).
}
\label{fig:improved_examples}
\vspace{-3mm}
\end{figure*}

\subsection{Evaluation of Drum Transcription}

We evaluated the effectiveness 
 of the proposed regularized training method based on the pretrained LMs.
Our transcription model
 was inspired by the state-of-the-art ADT method~\cite{vogl2017drum} (Fig.~\ref{fig:acoustic_model}).
The encoder consisted of 4 convolutional layers
 with the kernel size of $3\times3$ 
 and the decoder consisted of 3 GRU layers with 98 hidden dimensions,
 followed by a drop-out layer ($p=0.3$).
The weighting factor $\gamma$ in the transcription loss (Eq.~(\ref{sec:l_tran}))
 was set to $\gamma=0.46$ for the bi-gram model and $\gamma=0.61$ for the GRU model.
The weighting factor $\alpha$ in the total loss (Eq.~(\ref{sec:l_total}))
 was set to $\alpha=0.068$ for the bi-gram model and $\alpha=0.055$ for the GRU model.
The influential hyperparameters, 
 \ie, the number of GRU layers, the hidden dimension, $\gamma$, and $\alpha$
 were optimized for the validation data with Optuna\cite{optuna_2019}.
The weights of the convolutional and GRU layers were initialized based on \cite{kaiming2015delving},
 the fully connected layer was initialized by the sampling from ${\rm Uniform}(0,1)$,
 and the biases were initialized to 0.
We used AdamW optimizer~\cite{Loshchilov2019decoupled}
 with the initial learning rate of $10^{-3}$,
 the weight decay of $\lambda{=}10^{-4}$,
 $\beta_1=0.9$, $\beta_2=0.999$, and $\epsilon=10^{-9}$.
The threshold for $\bm\phi$ was set to $\delta=0.2$.
 
For comparison,
 we tested the state-of-the-art purely \textit{frame-level} ADT method\cite{vogl2017drum}
 based on a CRNN whose architecture was similar to our transcription model.
This model was trained with the following frame-level cross entropy:
\begin{align}
&\mathcal{L}_{\mathrm{tran}^*}(\bm\phi^*|\hat\mY^*) 
\nonumber\\
&=
- \sum_{k=1}^{K} \sum_{t=1}^{T}
\bigl(\beta \hat{Y}^*_{k, t}\log \phi^*_{k, t} 
+
(1 - \hat{Y}^*_{k, t}) \log (1 - \phi^*_{k, t}) \bigr),
\end{align}
where
 $\bm\phi^*, \hat\mY^* \in {\mathbb R}^{K \times T}$
 are the estimated onset probabilities and the ground-truth binary activations, respectively,
 and $\beta > 0$ is a weighting factor ($\beta=8$ in this paper). 
For each drum $k$, a frame $t$ was picked as an onset if
\begin{flalign*}
\text{1.\quad}& \phi^* = \max \{ \phi^*_{k, t-w_1:t+w_2}  \}, &\\
\text{2.\quad}& \phi^* \geq \mathrm{mean} \{ \phi^*_{k, t-w_3:t+w_4}  \} + \hat{\delta}, &\\
\text{3.\quad}& t - t_{\mathrm{prev}} > w_5,
\end{flalign*}
where $\hat{\delta}$ was a threshold,
 $w_{1:5}$ were interval parameters,
 and $t_{\mathrm{prev}}$ was the previous onset frame,
 which were set to $\hat{\delta}=0.2$,
 $w_1=w_3=w_5=2$,
 and $w_2=w_4=0$
 as in \cite{vogl2017drum}.
To measure the tatum-level transcription performance,
 the estimated frame-level onset times
 were quantized at the tatum level
 referring to the estimated or ground-truth tatum times.

Table~\ref{table:proposed_method_measures}
 shows the performances of the conventional frame-to-frame method~\cite{vogl2017drum} 
 followed by the frame-to-tatum quantization (post-processing)
 and the proposed frame-to-tatum method
 when the estimated or ground-truth beat times were given. 
We confirmed that the regularization method was 
 effective for improving the transcription model.
The regularization with the GRU model
 improved the F-measure by a larger margin
 than that with the bi-gram model.

Fig.~\ref{fig:improved_examples} illustrates
 two examples of transcribed drum scores,
 which show the positive effect of the language model-based regularization.
In both examples,
 the non-regularized transcription model
 often yielded musically-unnatural drum patterns,
 while the regularized model effectively avoided such patterns.
The regularized model, however,
 yielded extra onset times of hi-hats
 because the other kinds of percussive instruments (crash cymbals in both cases)
 were used instead of hi-hats.
We also found that
 the regularization mechanism was effective to estimate regular drum patterns,
 but tended to oversimplify highly-sophisticated non-regular drum patterns (\eg, fill-ins).

\begin{table}[t]
\centering
\caption{
The drum transcription performances of the conventional and proposed methods ($\%$).
}
\vspace{-2mm}
\begin{tabular}{l|cccccc} \toprule
& \multicolumn{3}{c}{Madmom} & \multicolumn{3}{c}{Ground-truth} \\
& $\mathcal{F}$ & $\mathcal{P}$ & $\mathcal{R}$ & $\mathcal{F}$ & $\mathcal{P}$ & $\mathcal{R}$ \\ \midrule
CRNN~\cite{vogl2017drum} & 70.8 & 77.4 & 65.9 & 71.0 & 77.6 & 66.1 \\
CRNN  & 78.9 & 86.3 & 73.1 & 79.3 & 86.7 & 73.3 \\
+ Bi-gram ($\alpha=0.068$) & 81.4 & 84.7 & 79.1 & 80.8 & 83.7 & 78.8 \\
+ GRU \quad\;($\alpha=0.055$) & {\bf 81.6} & 84.0 & 80.2 & {\bf 81.1} & 83.2 & 79.7 \\ \bottomrule
\end{tabular}
\label{table:proposed_method_measures}
\vspace{-3mm}
\end{table}

\section{Conclusion}\label{sec:conclusion}
This paper described a tatum-level ADT method 
 based on a CRNN trained with an LM-based regularization mechanism. 
This network consists of a frame-level convolutional encoder
 extracting the latent features of music signals
 and a tatum-level recurrent decoder
 considering musically-meaningful structure.
The experimental results showed that
 the regularized training significantly 
 improves both the correctness and musical naturalness of estimated drum scores.

Extending this approach, 
 we plan to deal with sophisticated and/or non-regular drum patterns (\eg, fill-ins)
 played by various kinds of percussive instruments (\eg, cymbals and toms). 
Considering that beat and downbeat times are closely related to drum patterns,
 it would be beneficial to integrate beat tracking into ADT
 in a multi-task learning framework. 

\section*{Acknowledgment}
This work is partially supported
 by JST ACCEL No. JPM-JAC1602
 and JSPS KAKENHI No. 16H01744, No. 19K20340, and No. 19H04137.

\bibliographystyle{unsrt}
\bibliography{main}

\end{document}